\documentclass{aa}

\usepackage{txfonts}
\usepackage{graphicx}
\usepackage{natbib}
\bibpunct{(}{)}{;}{a}{}{,}

\newcommand{\sdss}{\emph{SDSS }}

\newcommand{\teff}{$T_{\rm eff}$}
\newcommand{\logg}{$\log{g}$}

\newcommand{\kms}{$\rm{km~s^{-1}}$}
\newcommand{\masyr}{$\rm{mas~yr^{-1}}$}

\newcommand{\M}{$[M/H]$}

\begin{document}


\hyphenation{ana-lysed do-mi-nant}
\title{SDSS~J013655.91+242546.0  - an A-type hyper-velocity star from the outskirts of the
Galaxy
\thanks{Based on data collected at the European Southern Observatory, Chile.
Program ID: 082.D-0649}
\thanks{Based on observations at the 
3.5\,m telescope at DSAZ observatory (Calar Alto) in Spain. Program ID:
H09-3.5-028}}
\author{A. Tillich\inst{1}\and N. Przybilla\inst{1}\and R.-D. Scholz\inst{2}
\and U.  Heber\inst{1}
}
\offprints{A.~Tillich
\email{Alfred.Tillich@sternwarte.uni-erlangen.de}}
\institute{Dr. Remeis-Sternwarte Bamberg, Universit\"at Erlangen-N\"urnberg, 
Sternwartstr. 7, D-96049 Bamberg, Germany\and Astrophysikalisches Institut Potsdam, An der Sternwarte 16, 
D-14482 Potsdam}

\date{Received / Accepted }

\abstract{
Hyper-velocity stars (HVS) are moving so fast that they are unbound to the 
Galaxy. Dynamical ejection 
by a supermassive black hole is favoured to explain 
their origin. 
}
{
Locating the place of birth of an individual HVS is of utmost
importance to understanding the ejection mechanism.
}
{
SDSS~J013655.91+242546.0 (J0136+2425 for short) was found 
amongst three high-velocity 
stars (drawn from a sample of more than 10 000 blue stars), for which 
proper motions were measured. A kinematical as well as 
a quantitative 
NLTE spectral analysis was performed.  
When combined with the radial velocity (RV) and 
the spectroscopic distance, the trajectory of the star in
the Galactic potential was reconstructed. 
}
{
J0136+2425 is found to be an A-type 
main-sequence star travelling at $\approx$590~\kms, possibly unbound to the 
Galaxy and originating in the outer Galactic rim nowhere near the Galactic 
centre.
}
{
 J0136+2425 is the 
second HVS candidate with measured proper motion, besides the massive B star
HD~271791,
 and also the second 
for which its proper motion excludes a Galactic centre 
origin and, hence, the SMBH slingshot mechanism. Most known HVS are late B-type 
stars of about 3 M$_\odot$. With a mass of 2.45 M$_\odot$, J0136+2425 
resembles a typical HVS far more 
than HD~271791 does. Hence, this is the first time that a typical HVS is found 
not to originate in the Galactic centre. 
Its ejection velocity from the disk is so high (550~\kms) that the
extreme supernova binary scenario proposed for HD~271791 
is very unlikely. 
}


\titlerunning{J0136+2425 - an A-type hyper-velocity star from the outskirts of the
Galaxy}

\maketitle

\section{Introduction}\label{sec:intro}
Stars travelling so fast that they escape from the Galaxy are an inevitable 
consequence of the presence of a supermassive black 
hole (SMBH) in a dense stellar environment \citep{1988Natur.331..687H} such as
the Galactic centre (GC). \citet{1988Natur.331..687H} coined the term hyper-velocity
star (HVS) for such an object. It took a long time until 
 \cite{2005ApJ...622L..33B}, \cite{2005A&A...444L..61H} and 
\cite{2005ApJ...634L.181E} discovered the first three 
HVSs serendipitously and the interest in these stars grew tremendously.
Systematic searches for HVSs took advantage of the huge \sdss 
database \citep{2008ApJS..175..297A} for target selection 
and revealed a population of HVSs; the
latest compilation lists 16 of these stars \citep{2009ApJ...690.1639B}. 
These surveys are based on RV measurements alone. The known HVSs are 
non-uniformly distributed on the sky, as are their travel times.
\citet{2009ApJ...690L..69B} argue that this anisotropy supports their common origin being 
the Galactic
centre, while \citet{2009ApJ...691L..63A} point out that the overdensity of HVSs in the
constellation Leo suggests that the anisotropic distribution and 
preferred travel time are the result of the tidal disruption of a dwarf galaxy
in the Galactic potential.
          
Another issue is the distance of a HVS star, because blue horizontal branch (BHB) 
stars can not be easily distinguished from main-sequence (MS) stars for the HVSs in
question since both types of
stars populate the same region
in the \teff-$\log{g}$-diagram \citep{2008ASPC..392..167H}, 
but have different distances. It is generally assumed that the stars are main-sequence 
and not blue horizontal branch stars. Detailed spectroscopic analyses 
have confirmed these assumptions for HVS1, HVS3, HVS7, and HVS8 
\citep{2008A&A...480L..37P,2008A&A...488L..51P,2008ASPC..392..167H,2008ApJ...685L..47L}.  
In the absence of proper motion measurements, 
the trajectories have not been derived for any individual HVS so far.
\citet{2008A&A...483L..21H} succeeded in investigating the trajectory of
the high-velocity B star HD~271791 from accurate proper motions, radial velocity, 
and spectroscopic distance. The star was found to be probably unbound to the
Galaxy and to originate in the outer rim of the Galaxy rather than in its centre,
demonstrating that a mechanism other than that of \citet{1988Natur.331..687H}
must operate. Hence, it is rewarding to measure proper motions of
high-velocity stars. \citet{2008ApJ...684.1143X} presented radial velocities 
for more than 10 000 blue stars from the \sdss, which are a mix of blue horizontal
branch, blue straggler, and main-sequence stars with effective temperature
between roughly 7 000 and 10 000\,K according to their colours.
We focused on the 11 fastest stars in terms of large positive Galactic 
rest-frame (GRF) velocities to unravel their nature, distance, and
kinematics from detailed quantitative spectral analyses and astrometry.
Three stars exhibited significant proper motions\footnote{Two of the stars are found to
be a metal-poor straggler of population II and a spectroscopic binary,
as described in a forthcoming paper}.
Here we report that J0136+2425 is an A-type 
main sequence star travelling at $\approx$590~\kms, possibly unbound to the 
Galaxy and originating in the outer Galactic disk.
 




\section{Target selection and proper motion}\label{sec:tar+pm}

\begin{figure}[t]
\begin{center}
\includegraphics[scale=0.35]{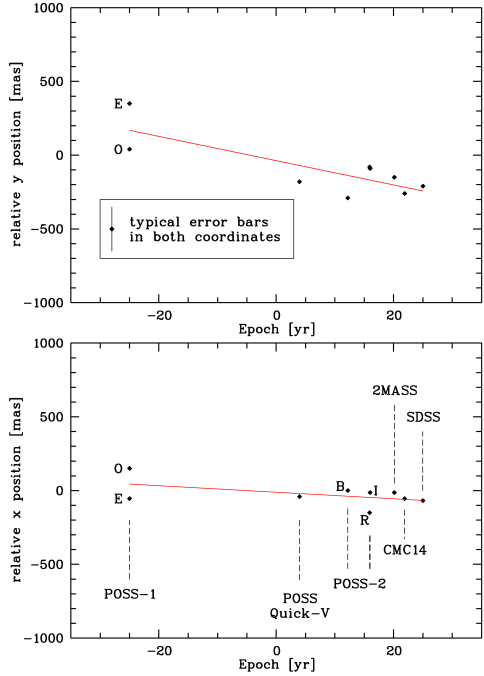}
\caption{\label{fig_PMfitJ01}Linear fit of the position measurements for J0136+2425, whereas 1979.74 is the zero epoch.}
\end{center}
\end{figure}

We selected all stars 
with GRF velocities $v_{\rm GRF}>+350$~\kms\ from the 
RV-based sample of \cite{2008ApJ...684.1143X} and obtained 11 
targets for which we attempted to measure proper motions. 
All 
available independent position measurements on Schmidt plates
(APM - \cite{2000yCat.1267....0M};
SSS - \cite{2001MNRAS.326.1279H}) were collected
and combined with the SDSS and other available positions
(CMC14~\cite{2006yCat.1304....0C};
2MASS - \cite{2003tmc..book.....C};
UKIDSS - \cite{2007MNRAS.379.1599L})
to perform a linear proper motion fit. However, there were even
more Schmidt plate measurements from up to 14 different epochs in the case
of overlapping plates of the Digitised Sky
Surveys\footnote{http://archive.stsci.edu/cgi-bin/dss\_plate\_finder}.
FITS images of 15 by 15 arcmin size were extracted from all available plates
 and ESO MIDAS tool center/gauss was used to measure positions. For this purpose,
we selected compact background galaxies around each target, identified from
SDSS, to transform the target positions in all the Schmidt plates to the
SDSS system. The small fields allowed us to apply a simple model
(shift+rotation) and to achieve an improved fit
for all our targets (see Fig.~\ref{fig_PMfitJ01} ). 
Significant proper motions were found for the three brightest stars only.
For J0136+2425, we obtain $\mu_\alpha\cos(\delta)$ = $-$2.2 $\pm$ 1.3 \masyr\ and 
$\mu_\delta$  =  $-$8.2 $\pm$ 2.2~\masyr.

 
\section{Observations and quantitative spectroscopy}\label{sec:spec}

In order to exclude RV variability, we reobserved J0136+2425 at ESO 
with EFOSC2 mounted on the NTT in October 2008 and with  
 the TWIN spectrograph at the 3.5m telescope on Calar Alto in  
 July 2009; during the latter run, six spectra for J0136+2425 
 distributed over three days were
 obtained. Radial velocities were derived by 
 $\chi^2$-fitting suitable synthetic spectra over the full spectral range.
 Since we used many spectral lines, our results differ from that of
 \citet{2008ApJ...684.1143X} who used the $\mathrm H{\delta}$ line only.  
The radial velocities from individual spectra agree to within 
their respective error limits, indicating that the star is not RV
variable within a few kilometers per second on timescales of days.

\begin{figure*}
\begin{center}
\resizebox{.99\hsize}{!}{\includegraphics{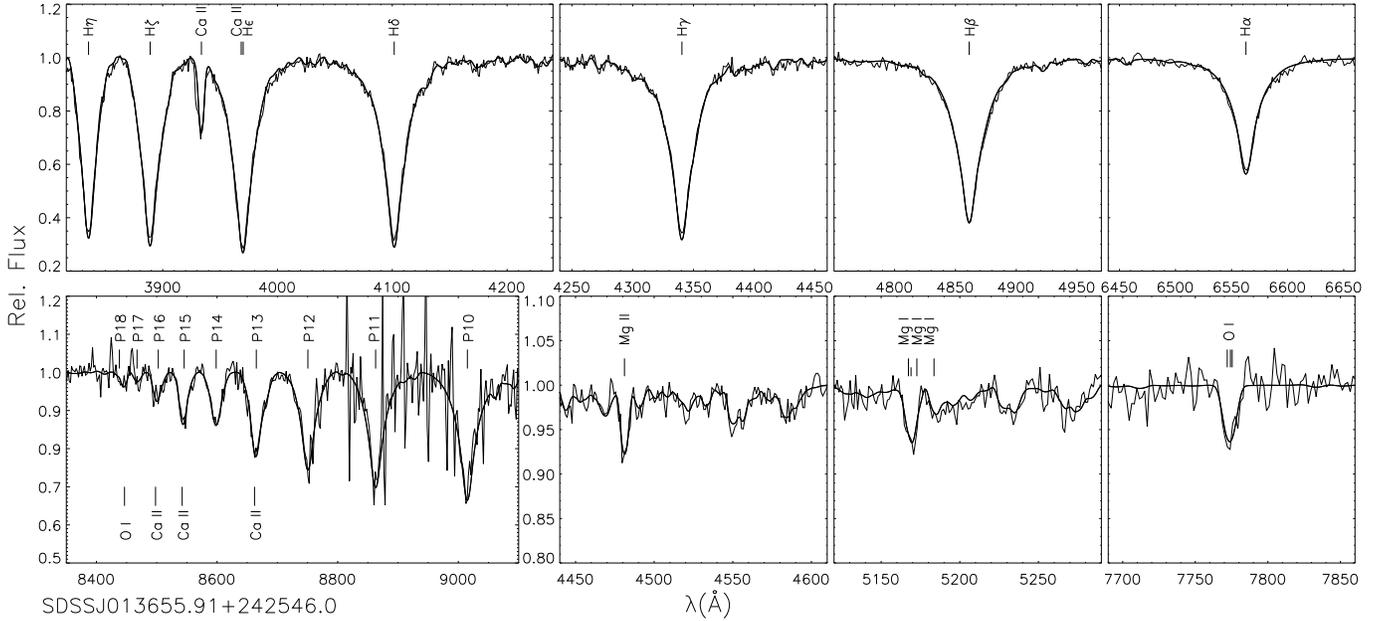}}\\[1.95mm]
\caption{\label{linefits}
Comparison of NLTE spectrum synthesis (thick line) with observation 
(thin wiggly line) for J0136+2425. 
}
\end{center}
\end{figure*}

A quantitative analysis 
was carried
out following the hybrid NLTE approach discussed by \cite{2006A&A...445.1099P}.
The effective temperature {\teff} and the surface gravity
$\log g$ were determined by fits to the Stark-broadened Balmer and Paschen
lines and the ionisation equilibrium of \ion{Mg}{i/ii}. The stellar
metallicity were derived by model fits to the observed metal-line
spectra. Results are listed in Table \ref{tab_HVS} and a
 comparison of the resulting final synthetic spectrum with
observations in the selected
regions around the Balmer lines, the higher Paschen series,
\ion{Mg}{ii}\,$\lambda$4481{\AA}, the \ion{Mg}{i}\,b and the near-IR
\ion{O}{i} triplets, is shown in Fig.~\ref{linefits}.
Overall, excellent agreement is obtained for all strategic spectral lines
throughout the entire wavelength range. 
The uncertainties in the stellar parameters were constrained by
the quality of the match of the spectral indicators within the given S/N
limitations. 

\begin{figure}
\begin{center}
\includegraphics[scale=0.7]{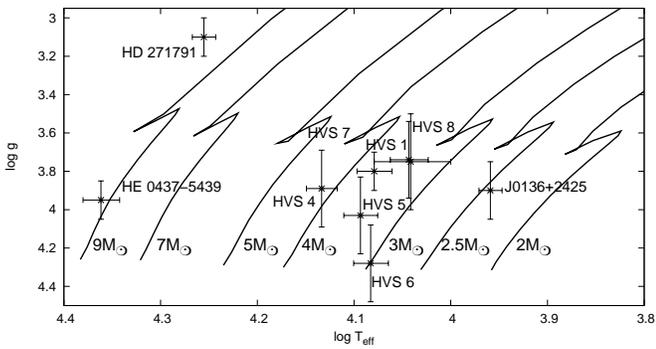}
\caption{\label{teff_g_j0136}
J0136+2425 in the (\teff,\logg) diagram with evolutionary 
tracks of \citet{1992A&AS...96..269S} for solar metallicity. 
}
\end{center}
\end{figure}

Its \teff\ and gravity places J0136+2425 on the main sequence (see Fig. 
\ref{teff_g_j0136}). In
addition the star is rapidly rotating at 250~\kms\ and its metallicity is solar,
 which strengthens the
conclusion that it is an intermediate-mass A-type main-sequence star of 2.45~M$_\odot$ as
derived by comparing the position of the star to predictions of the
evolutionary models of \cite{1992A&AS...96..269S}.

In Fig.~\ref{teff_g_j0136}, we compare the position of J0136+2425 in the 
(\teff, \logg)-diagram to those of five late B-type HVSs for which these
parameters are available \citep{2008A&A...480L..37P,2008A&A...488L..51P,2008ASPC..392..167H,2008ApJ...685L..47L} as well as those of 
the massive B
star HE~0437$-$5439 originating in the LMC, and the B giant HD~271791. The
vast majority of HVSs are of late B type because they were discovered by
targeted searches. Hence, we shall term them typical HVSs. 
J0136+2425 is slightly cooler and less massive than the other known HVS. However,
its mass and evolutionary lifetime is similar to that of the typical HVSs
(3--4 M$_\odot$, $\approx$100~Myrs), while HE~0437$-$5439 and HD~271791 are far 
more massive (9--11~M$_\odot$) and short-lived ($\approx$20~Myrs).
 
\begin{table}
\caption{Results of the spectroscopic and kinematic analysis of J0136+2425. 
}
\label{tab_HVS}
\begin{center}
\begin{tabular}{lc|lc}
\hline\hline
$V$ (mag)$^{\rm a} $     & 16.17 $\pm$ 0.02 & 
$E(B-V)$ (mag)$^{\rm b} $ &  0.16 $\pm$ 0.02 \\
$\mu_\alpha\cos(\delta)$\,(\masyr) &   $-$2.2 $\pm$ 1.3 & 
$\mu_\delta$\,(\masyr) &   $-$8.2 $\pm$ 2.2 \\
\teff\ (K)      &  9100 $\pm$ 250  & 
 \logg (cgs)      &  3.90 $\pm$ 0.15  \\
 \M          &  0.0             &  
 $M/M_\odot$ &  2.45 $\pm$ 0.20   \\
 $v_{\rm rad}$ (\kms)   & 324.3 $\pm$ 5.9 &
 $v_{\rm rot} \sin i$ (\kms) & 250 \\ 
 $d$ (kpc)    & 10.90 $\pm$ 2.00  &    
 $v_{\rm grf}$ (\kms) & 587 $^{+144}_{-89}$   \\
 $v_{\rm ej}$ (\kms)  & 551                 & 
 $v_{\rm esc}$ (\kms) & 466                  \\
 $TOF$ (Myrs)           &  12 $\pm$ 1.3  & t$_{\rm evol}$ (Myrs) & 245    \\
 \hline  
 \multicolumn{4}{l}{\small $^{\rm a}$ The visual magnitude 
derived following \cite{2006A&A...460..339J}}\\
 \multicolumn{4}{l}{\small $^{\rm b}$ The interstellar colour excess
$E(B-V)$ was determined}\\
 \multicolumn{4}{l}{\small 
 by comparing the
observed colours to synthetic ones}\\
 \multicolumn{4}{l}{\small 
 from the model spectral energy
distribution. }\\

\end{tabular}
\end{center}
\end{table}

\section{Distance, kinematics, and errors}\label{sec:kine}

Using the mass, effective temperature, gravity, and extinction-corrected apparent
magnitude, we derive the distance following \cite{2001A&A...378..907R} 
using the fluxes from
the final model spectrum. The distance error is dominated by the gravity error. 

\begin{figure}
\begin{center}
\includegraphics[scale=0.65]{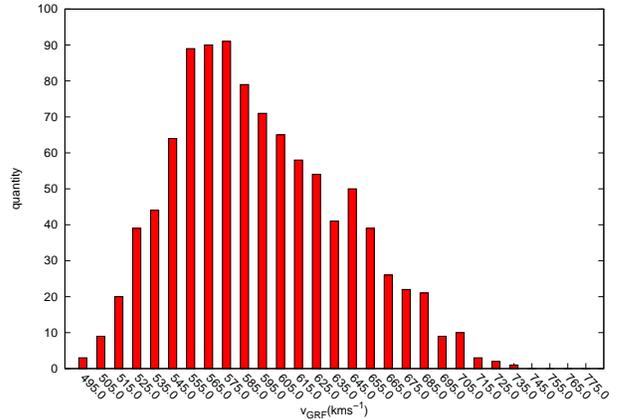}
\caption{\label{fig:veldistrib_J0136}   
Galactic rest-frame velocity distribution for J0136+2425. 
}
\end{center}
\end{figure}

Applying the Galactic potential of
\citet{1991RMxAA..22..255A}, we calculated orbits and 
reconstructed the path of the star back to the Galactic plane with the program of
\citet{1992AN....313...69O}. 
The distance of the GC from the Sun was adopted to be 8.0~kpc and
the Sun's motion with respect to the local standard of rest was taken from 
\citet{1998MNRAS.298..387D}. 
Since the RV is well known, the error in the space motion is made up of 
that of the distance, which is controlled mainly by the gravity error, and those of the 
proper
motion components. Varying these three quantities within their respective errors 
by applying a Monte Carlo procedure with a depth of 1000, we determined the intersection area of the
trajectories with the Galactic plane and the 
time-of-flight. From these Monte Carlo simulations, we derived 
the median GRF velocities at the present location and their distibution 
(see Fig.~\ref{fig:veldistrib_J0136}) and
compared these with the local escape velocity calculated from the Galactic potential
of \citet{1991RMxAA..22..255A}. 

\begin{figure}
\includegraphics[scale=0.7]{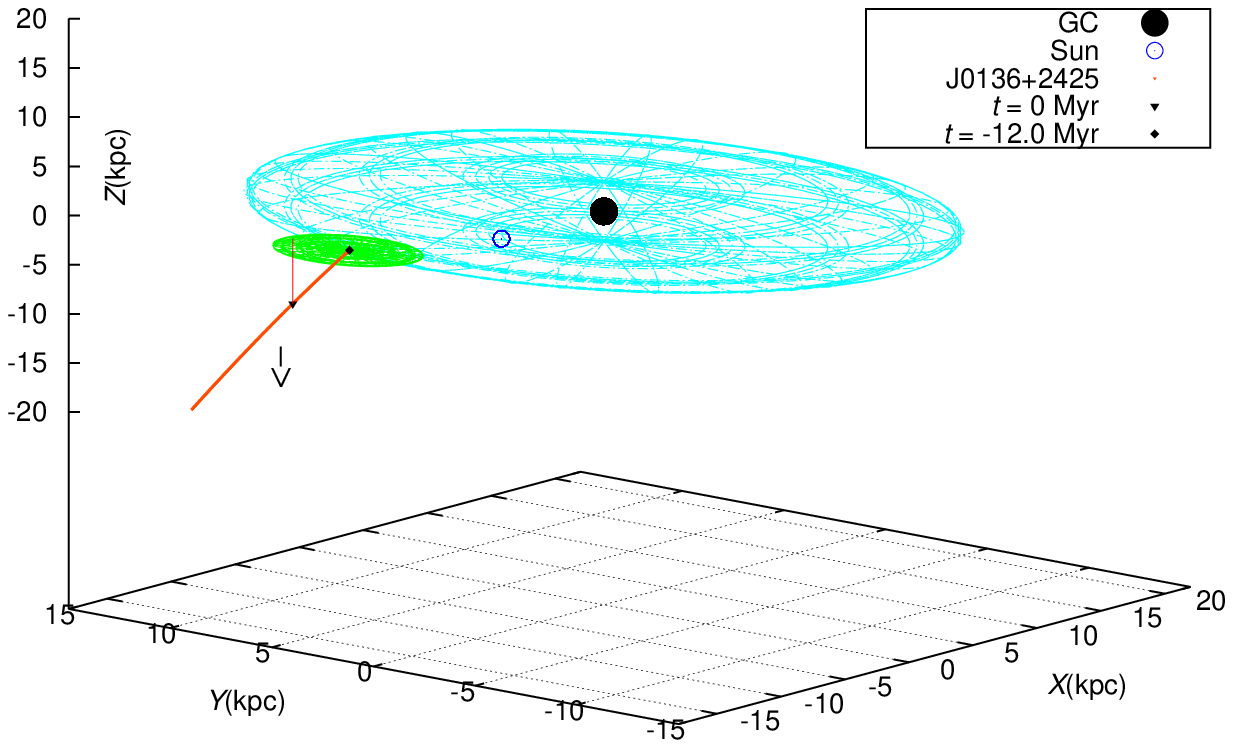}
\includegraphics[scale=0.7]{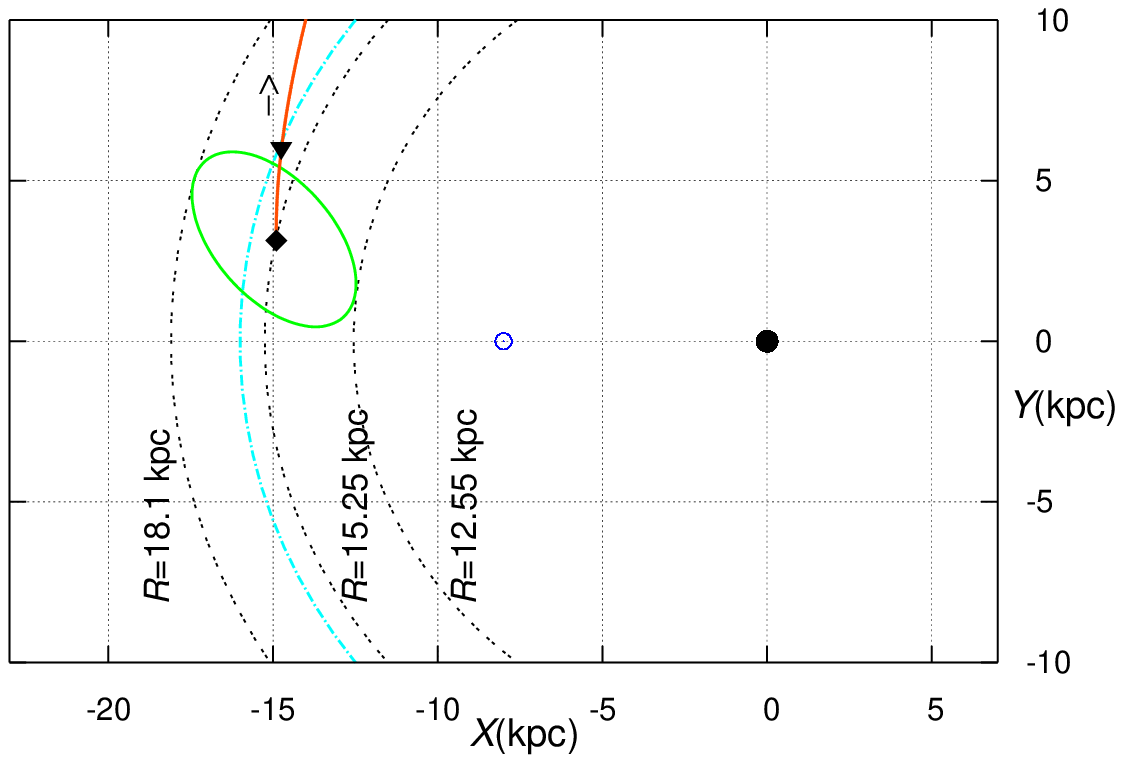}

\caption{\label{fig_J0136_2d3dplot}{\bf Upper panel:}
Trajectory for J0136+2425 with place of birth marked in green (3$\sigma$-error), relatively
 to the Galactic disk (light blue). 
{\bf Lower panel:} Galactic disk (light blue) projection of the 
trajectory for J0136+2425 and intersection of the bunch of trajectories (green). 
Note that the GC is far from the plane intersection area.
}
\end{figure}

%
For J0136+2425, the GRF velocity of 587$^{+144}_{-89}$~\kms\ 
slightly exceeds the local
escape velocity. Whether the star is bound to the Galaxy depends on the Galactic
potential adopted, in particular for the dark matter halo, as pointed out by 
\citet{2009ApJ...691L..63A}. \citet{1991RMxAA..22..255A} adopted a halo mass out
to 100~kpc of
$M_{\rm Halo}$= 8$\times$10$^{11}$M$_\odot$. \citet{2008ApJ...684.1143X} 
derived a
somewhat lower mass 
whereas  
\citet{2009ApJ...691L..63A} favoured a higher one of 
$M_{\rm Halo}$= 1.83$\times$10$^{12}$~M$_\odot$. If the former were correct, 
J0136+2425 would be unbound, while it would be bound in the latter case.
As can be seen in Fig. \ref{fig_J0136_2d3dplot}, 
its place of origin 
is found to be in the outer part of the Galactic plane 
at Galactic radii between 12.5~kpc and 18~kpc, nowhere near the GC. 
The time of flight (12 $\pm$ 1.3~Myr) is much shorter than the star's lifetime (450~Myr).

%

\section{Conclusion}\label{sec:conclusion} 

We have reported the quantitative spectral analysis of a high-velocity star from 
the sample of faint blue stars in the halo of \citet{2008ApJ...684.1143X}.
The radial velocity, proper motion, and spectroscopic distance were derived and
a detailed kinematical analysis was performed using the Galactic potential 
of \citet{1991RMxAA..22..255A}.  

%
J0136+2425 was found to be a rapidly rotating A star of solar composition and
therefore classified as a main-sequence star of 2.45~M$_\odot$. The kinematic
analysis 
excludes an origin at the GC and hence the Hills mechanism for
ejection of the star. 
Its place of origin
was found to be in the outer part of the Galactic plane 
at Galactic radii between 12.5~kpc and 18~kpc. This is very similar to the case 
of the massive B star HD~271791 \citep{2008A&A...483L..21H}, which was the first 
ultra-high-velocity star whose proper motion excludes a GC origin. 
The ejection mechanism for the star was proposed to be an extreme binary supernova, 
which also explained its enrichment in $\alpha$ elements. 
However, we find no evidence of $\alpha$-enhancement in J0136+2425 from the
existing spectra. The star is more metal-rich than expected on average for an
object born at such a large Galactocentric distance. Due to the presence of
Galactic abundance gradients \citep{2006ApJS..162..346R}, we expect the outer parts of 
the Galaxy to be less metal-rich than the Sun. 
The metallicity of J0136+2425 is compatible with the high end of the
abundance distribution in the outer disk.
 
The required ejection velocity from the disk is so high 
(550~\kms) that an extreme supernova binary scenario as proposed for HD~271791 
\citep{2008ApJ...684L.103P} is very unlikely. 

Most known HVS are late B-type 
stars of about 3 M$_\odot$. With a mass of 2.45 M$_\odot$, J0136+2425 
resembles such a typical HVS much more 
than HD~271791 ($M=$11 M$_\odot$) does. 
Hence, this is the first time that a typical HVS is found 
not to originate in the GC and excludes the SMBH slingshot
mechanism. Hence, typical HVS may have been ejected by different 
mechanisms other than that proposed by Hills.    

Once more, this calls for an alternative ejection scenario to the Hills mechanism
 such as dynamical ejection from
clusters or binary supernovae \citep[see][]{2009MNRAS.396..570G}. 
Hence, we are left with the dynamical ejection scenario or tidal disruption 
of a satellite galaxy, as proposed by \cite{2009ApJ...691L..63A}. 
They noticed a clustering
of more than half of the known HVS in a region of 26$^\circ$-diameter in the 
constellation of Leo 
($l_{II}\approx 230^\circ$, $b_{II}\approx 60^\circ$). 
They suggested that if these star 
stems from a disruption event of a dwarf galaxy, more high velocity stars 
should be found in that area. However, J0136+2425 ($l_{II}=136^\circ$,
$b_{II}=-37^\circ$) is 
located far from Leo and is therefore probably unrelated.

\acknowledgements{A.T. acknowledges funding by the Deutsche Forschungsgemeinschaft through  grant HE1356/45-1. 
We are very grateful to Stephan Geier for stimulating discussions and advice.
Our thanks go to S. M\"{u}ller and T. Kupfer for observing and reducing the 
data from DSAZ.}

\bibliography{references}

\begin{thebibliography}{29}
\expandafter\ifx\csname natexlab\endcsname\relax\def\natexlab#1{#1}\fi

\bibitem[{{Abadi} {et~al.}(2009){Abadi}, {Navarro}, \&
  {Steinmetz}}]{2009ApJ...691L..63A}
{Abadi}, M.~G., {Navarro}, J.~F., \& {Steinmetz}, M. 2009, \apjl, 691, L63

\bibitem[{{Adelman-McCarthy} {et~al.}(2008){Adelman-McCarthy}, {Ag{\"u}eros},
  {Allam}, {Allende Prieto}, {Anderson}, {Anderson}, {Annis}, {Bahcall},
  {Bailer-Jones}, {Baldry}, {Barentine}, {Bassett}, {Becker}, {Beers}, {Bell},
  {Berlind}, {Bernardi}, {Blanton}, {Bochanski}, {Boroski}, {Brinchmann},
  {Brinkmann}, {Brunner}, {Budav{\'a}ri}, {Carliles}, {Carr}, {Castander},
  {Cinabro}, {Cool}, {Covey}, {Csabai}, {Cunha}, {Davenport}, {Dilday}, {Doi},
  {Eisenstein}, {Evans}, {Fan}, {Finkbeiner}, {Friedman}, {Frieman},
  {Fukugita}, {G{\"a}nsicke}, {Gates}, {Gillespie}, {Glazebrook}, {Gray},
  {Grebel}, {Gunn}, {Gurbani}, {Hall}, {Harding}, {Harvanek}, {Hawley},
  {Hayes}, {Heckman}, {Hendry}, {Hindsley}, {Hirata}, {Hogan}, {Hogg}, {Hyde},
  {Ichikawa}, {Ivezi{\'c}}, {Jester}, {Johnson}, {Jorgensen}, {Juri{\'c}},
  {Kent}, {Kessler}, {Kleinman}, {Knapp}, {Kron}, {Krzesinski}, {Kuropatkin},
  {Lamb}, {Lampeitl}, {Lebedeva}, {Lee}, {Leger}, {L{\'e}pine}, {Lima}, {Lin},
  {Long}, {Loomis}, {Loveday}, {Lupton}, {Malanushenko}, {Malanushenko},
  {Mandelbaum}, {Margon}, {Marriner}, {Mart{\'{\i}}nez-Delgado}, {Matsubara},
  {McGehee}, {McKay}, {Meiksin}, {Morrison}, {Munn}, {Nakajima}, {Neilsen},
  {Newberg}, {Nichol}, {Nicinski}, {Nieto-Santisteban}, {Nitta}, {Okamura},
  {Owen}, {Oyaizu}, {Padmanabhan}, {Pan}, {Park}, {Peoples}, {Pier}, {Pope},
  {Purger}, {Raddick}, {Re Fiorentin}, {Richards}, {Richmond}, {Riess}, {Rix},
  {Rockosi}, {Sako}, {Schlegel}, {Schneider}, {Schreiber}, {Schwope}, {Seljak},
  {Sesar}, {Sheldon}, {Shimasaku}, {Sivarani}, {Smith}, {Snedden}, {Steinmetz},
  {Strauss}, {SubbaRao}, {Suto}, {Szalay}, {Szapudi}, {Szkody}, {Tegmark},
  {Thakar}, {Tremonti}, {Tucker}, {Uomoto}, {Vanden Berk}, {Vandenberg},
  {Vidrih}, {Vogeley}, {Voges}, {Vogt}, {Wadadekar}, {Weinberg}, {West},
  {White}, {Wilhite}, {Yanny}, {Yocum}, {York}, {Zehavi}, \&
  {Zucker}}]{2008ApJS..175..297A}
{Adelman-McCarthy}, J.~K., {Ag{\"u}eros}, M.~A., {Allam}, S.~S., {et~al.} 2008,
  \apjs, 175, 297

\bibitem[{{Allen} \& {Santillan}(1991)}]{1991RMxAA..22..255A}
{Allen}, C. \& {Santillan}, A. 1991, Rev. Mex. Astr. Astrofis., 22, 255

\bibitem[{{Brown} {et~al.}(2009{\natexlab{a}}){Brown}, {Geller}, \&
  {Kenyon}}]{2009ApJ...690.1639B}
{Brown}, W.~R., {Geller}, M.~J., \& {Kenyon}, S.~J. 2009{\natexlab{a}}, \apj,
  690, 1639

\bibitem[{{Brown} {et~al.}(2009{\natexlab{b}}){Brown}, {Geller}, {Kenyon}, \&
  {Bromley}}]{2009ApJ...690L..69B}
{Brown}, W.~R., {Geller}, M.~J., {Kenyon}, S.~J., \& {Bromley}, B.~C.
  2009{\natexlab{b}}, \apjl, 690, L69

\bibitem[{{Brown} {et~al.}(2005){Brown}, {Geller}, {Kenyon}, \&
  {Kurtz}}]{2005ApJ...622L..33B}
{Brown}, W.~R., {Geller}, M.~J., {Kenyon}, S.~J., \& {Kurtz}, M.~J. 2005,
  \apjl, 622, L33

\bibitem[{Carlsberg-Meridian-Catalog(2006)}]{2006yCat.1304....0C}
Carlsberg-Meridian-Catalog. 2006, Copenhagen Univ. Obs., Inst. of Astr.,
  Cambridge, UK, Real\,Inst.\,y\,Obs.\,de\,la\,Armada\,en\,San\,Fernando, 1304,
  0

\bibitem[{{Cutri} {et~al.}(2003){Cutri}, {Skrutskie}, {van Dyk}, {Beichman},
  {Carpenter}, {Chester}, {Cambresy}, {Evans}, {Fowler}, {Gizis}, {Howard},
  {Huchra}, {Jarrett}, {Kopan}, {Kirkpatrick}, {Light}, {Marsh}, {McCallon},
  {Schneider}, {Stiening}, {Sykes}, {Weinberg}, {Wheaton}, {Wheelock}, \&
  {Zacarias}}]{2003tmc..book.....C}
{Cutri}, R.~M., {Skrutskie}, M.~F., {van Dyk}, S., {et~al.} 2003, {2MASS All
  Sky Catalog of point sources.}

\bibitem[{{Dehnen} \& {Binney}(1998)}]{1998MNRAS.298..387D}
{Dehnen}, W. \& {Binney}, J.~J. 1998, \mnras, 298, 387

\bibitem[{{Edelmann} {et~al.}(2005){Edelmann}, {Napiwotzki}, {Heber},
  {Christlieb}, \& {Reimers}}]{2005ApJ...634L.181E}
{Edelmann}, H., {Napiwotzki}, R., {Heber}, U., {Christlieb}, N., \& {Reimers},
  D. 2005, \apjl, 634, L181

\bibitem[{{Gvaramadze} {et~al.}(2009){Gvaramadze}, {Gualandris}, \& {Portegies
  Zwart}}]{2009MNRAS.396..570G}
{Gvaramadze}, V.~V., {Gualandris}, A., \& {Portegies Zwart}, S. 2009, \mnras,
  396, 570

\bibitem[{{Hambly} {et~al.}(2001){Hambly}, {MacGillivray}, {Read}, {Tritton},
  {Thomson}, {Kelly}, {Morgan}, {Smith}, {Driver}, {Williamson}, {Parker},
  {Hawkins}, {Williams}, \& {Lawrence}}]{2001MNRAS.326.1279H}
{Hambly}, N.~C., {MacGillivray}, H.~T., {Read}, M.~A., {et~al.} 2001, \mnras,
  326, 1279

\bibitem[{{Heber} {et~al.}(2008{\natexlab{a}}){Heber}, {Edelmann},
  {Napiwotzki}, {Altmann}, \& {Scholz}}]{2008A&A...483L..21H}
{Heber}, U., {Edelmann}, H., {Napiwotzki}, R., {Altmann}, M., \& {Scholz},
  R.-D. 2008{\natexlab{a}}, \aap, 483, L21

\bibitem[{{Heber} {et~al.}(2008{\natexlab{b}}){Heber}, {Hirsch}, {Edelmann},
  {Napiwotzki}, {O'Toole}, {Brown}, \& {Altmann}}]{2008ASPC..392..167H}
{Heber}, U., {Hirsch}, H.~A., {Edelmann}, H., {et~al.} 2008{\natexlab{b}}, in
  ASPCS, Vol. 392, Hot Subdwarf Stars and Related Objects, ed. U.~{Heber},
  C.~S. {Jeffery}, \& R.~{Napiwotzki}, 167

\bibitem[{{Hills}(1988)}]{1988Natur.331..687H}
{Hills}, J.~G. 1988, \nat, 331, 687

\bibitem[{{Hirsch} {et~al.}(2005){Hirsch}, {Heber}, {O'Toole}, \&
  {Bresolin}}]{2005A&A...444L..61H}
{Hirsch}, H.~A., {Heber}, U., {O'Toole}, S.~J., \& {Bresolin}, F. 2005, \aap,
  444, L61

\bibitem[{{Jordi} {et~al.}(2006){Jordi}, {Grebel}, \&
  {Ammon}}]{2006A&A...460..339J}
{Jordi}, K., {Grebel}, E.~K., \& {Ammon}, K. 2006, \aap, 460, 339

\bibitem[{{Lawrence} {et~al.}(2007){Lawrence}, {Warren}, {Almaini}, {Edge},
  {Hambly}, {Jameson}, {Lucas}, {Casali}, {Adamson}, {Dye}, {Emerson},
  {Foucaud}, {Hewett}, {Hirst}, {Hodgkin}, {Irwin}, {Lodieu}, {McMahon},
  {Simpson}, {Smail}, {Mortlock}, \& {Folger}}]{2007MNRAS.379.1599L}
{Lawrence}, A., {Warren}, S.~J., {Almaini}, O., {et~al.} 2007, \mnras, 379,
  1599

\bibitem[{{L{\'o}pez-Morales} \& {Bonanos}(2008)}]{2008ApJ...685L..47L}
{L{\'o}pez-Morales}, M. \& {Bonanos}, A.~Z. 2008, \apjl, 685, L47

\bibitem[{{McMahon} {et~al.}(2000){McMahon}, {Irwin}, \&
  {Maddox}}]{2000yCat.1267....0M}
{McMahon}, R.~G., {Irwin}, M.~J., \& {Maddox}, S.~J. 2000, VizieR Online Data
  Catalog, 1267, 0

\bibitem[{{Odenkirchen} \& {Brosche}(1992)}]{1992AN....313...69O}
{Odenkirchen}, M. \& {Brosche}, P. 1992, Astronomische Nachrichten, 313, 69

\bibitem[{{Przybilla} {et~al.}(2006){Przybilla}, {Butler}, {Becker}, \&
  {Kudritzki}}]{2006A&A...445.1099P}
{Przybilla}, N., {Butler}, K., {Becker}, S.~R., \& {Kudritzki}, R.~P. 2006,
  \aap, 445, 1099

\bibitem[{{Przybilla} {et~al.}(2008{\natexlab{a}}){Przybilla}, {Nieva},
  {Heber}, \& {Butler}}]{2008ApJ...684L.103P}
{Przybilla}, N., {Nieva}, M.~F., {Heber}, U., \& {Butler}, K.
  2008{\natexlab{a}}, \apjl, 684, L103

\bibitem[{{Przybilla} {et~al.}(2008{\natexlab{b}}){Przybilla}, {Nieva},
  {Heber}, {Firnstein}, {Butler}, {Napiwotzki}, \&
  {Edelmann}}]{2008A&A...480L..37P}
{Przybilla}, N., {Nieva}, M.~F., {Heber}, U., {et~al.} 2008{\natexlab{b}},
  \aap, 480, L37

\bibitem[{{Przybilla} {et~al.}(2008{\natexlab{c}}){Przybilla}, {Nieva},
  {Tillich}, {Heber}, {Butler}, \& {Brown}}]{2008A&A...488L..51P}
{Przybilla}, N., {Nieva}, M.~F., {Tillich}, A., {et~al.} 2008{\natexlab{c}},
  \aap, 488, L51

\bibitem[{{Ramspeck} {et~al.}(2001){Ramspeck}, {Heber}, \&
  {Moehler}}]{2001A&A...378..907R}
{Ramspeck}, M., {Heber}, U., \& {Moehler}, S. 2001, \aap, 378, 907

\bibitem[{{Rudolph} {et~al.}(2006){Rudolph}, {Fich}, {Bell}, {Norsen},
  {Simpson}, {Haas}, \& {Erickson}}]{2006ApJS..162..346R}
{Rudolph}, A.~L., {Fich}, M., {Bell}, G.~R., {et~al.} 2006, \apjs, 162, 346

\bibitem[{{Schaller} {et~al.}(1992){Schaller}, {Schaerer}, {Meynet}, \&
  {Maeder}}]{1992A&AS...96..269S}
{Schaller}, G., {Schaerer}, D., {Meynet}, G., \& {Maeder}, A. 1992, \aaps, 96,
  269

\bibitem[{{Xue} {et~al.}(2008){Xue}, {Rix}, {Zhao}, {Re Fiorentin}, {Naab},
  {Steinmetz}, {van den Bosch}, {Beers}, {Lee}, {Bell}, {Rockosi}, {Yanny},
  {Newberg}, {Wilhelm}, {Kang}, {Smith}, \& {Schneider}}]{2008ApJ...684.1143X}
{Xue}, X.~X., {Rix}, H.~W., {Zhao}, G., {et~al.} 2008, \apj, 684, 1143

\end{thebibliography}
\bibliographystyle{aa}
\end{document}